\documentclass[12pt]{article}

\usepackage{amsmath,amssymb,graphicx} %use drftcite in draftmode
\usepackage{pstricks}
\usepackage{bbm}
\usepackage{cite}
\usepackage{latexsym}
\newcommand{\centeron}[2]{{\setbox0=\hbox{#1}\setbox1=\hbox{#2}\ifdim
\wd1>\wd0\kern.5\wd1\kern-.5\wd0\fi \copy0
\kern-.5\wd0\kern-.5\wd1\copy1\ifdim\wd0>\wd1
                                   \kern.5\wd0\kern-.5\wd1\fi}}
\newcommand{\ltap}{\>\centeron{\raise.35ex\hbox{$<$}}
                           {\lower.65ex\hbox{$\sim$}}\>}
\newcommand{\gtap}{\>\centeron{\raise.35ex\hbox{$>$}}
                           {\lower.65ex\hbox{$\sim$}}\>}
\newcommand{\gsim}{\mathrel{\gtap}}

\newcommand\ZZ{\hbox{\zfont Z\kern-.4emZ}}
\font\zfont = cmss10 %scaled \magstep1
\newcommand{\sfrac}[2]{{\textstyle\frac{#1}{#2}}}

\newcommand{\fref}[1]{fig.\ \ref{f.#1}}
\newcommand{\eref}[1]{eq.\ (\ref{e.#1})}

\newcommand{\sref}[1]{Section \ref{s.#1}}
\newcommand{\ssref}[1]{Section \ref{ss.#1}}
\newcommand{\cref}[1]{Chapter \ref{c.#1}}
\newcommand{\tref}[1]{Table \ref{t.#1}}

\newcommand{\ba}{\begin{array}}
\newcommand{\ea}{\end{array}}

\newcommand{\beq}{\begin{eqnarray}}% can be used as {equation} or  {eqnarray}
\newcommand{\eeq}{\end{eqnarray}}
\newcommand{\beqs}{\begin{eqnarray*}}
\newcommand{\eeqs}{\end{eqnarray*}}

\newcommand{\bal}{\begin{align}} %this is glorious: arbitrary number of columns, alignment
\newcommand{\eal}{\end{align}}

\def\bi{\begin{itemize}}
\def\ei{\end{itemize}}
\def\ben{\begin{enumerate}}
\def\een{\end{enumerate}}
\def\bc{\begin{center}}
\def\ec{\end{center}}
\def\bt{\begin{table}}
\def\et{\end{table}}
\def\btb{\begin{tabular}}
\def\etb{\end{tabular}}

\def\gev{\, {\rm GeV}}
\def\tev{\, {\rm TeV}}

\def\mass2{mass${}^2$}

\newenvironment{tenumerate}{
\begin{enumerate}
  \setlength{\itemsep}{1pt}
  \setlength{\parskip}{0pt}
  \setlength{\parsep}{0pt}
  \vspace{-2mm}
}{\vspace{-2mm} \end{enumerate}}

\begin{document}
\bibliographystyle{unsrt}
\begin{titlepage}
%\begin{flushright}
%{\tt hep-ph/yymmnn}
%\end{flushright}

\vskip1.5cm
\begin{center}
{\huge \bf A Flavor Protection for Warped Higgsless Models}
\end{center}
\vskip0.2cm

\begin{center}
{\bf Csaba Cs\'aki and David Curtin}

\end{center}
\vskip 8pt

\begin{center}
{\it Institute for High Energy Phenomenology\\
Newman Laboratory of Elementary Particle Physics\\
Cornell University, Ithaca, NY 14853, USA } \\
\vspace*{0.3cm}

\vspace*{0.1cm}

{\tt csaki@cornell.edu, drc39@cornell.edu}
\end{center}

\vglue 0.3truecm

\begin{abstract}
We examine various possibilities for realistic 5D higgsless models and construct a full quark sector featuring next-to-minimal flavor violation (with an exact bulk $SU(2)$ protecting the first two generations) satisfying electroweak and flavor constraints. The ``new custodially protected representation'' is used for the third generation to protect the light quarks from flavor violations induced due to the heavy top. A combination of flavor symmetries, and RS-GIM for the right-handed quarks suppresses flavor-changing neutral currents below experimental bounds, assuming CKM-type mixing on the UV brane. In addition to the usual higgsless RS signals, this model predicts an exotic charge-$5/3$ quark with mass of about 0.5 TeV which should show up at the LHC very quickly, as well as nonzero flavor-changing neutral currents which could be detected in the next generation of flavor experiments. In the course of our analysis, we also find quantitative estimates for the errors of the fermion zero mode approximation, which are significant for higgsless-type models.
\end{abstract}

\end{titlepage}

%%%%%%%%%%%%%%%%%%%%%%%%%%%%%%%%%%%%%%%%%%%%%%%%%%%%%%%%%%%%%%%%%%%%%%
%%%%%%%%%%%%%%%%%%%%%%%%%%%%%%%%%%%%%%%%%%%%%%%%%%%%%%%%%%%%%%%%%%%%%%%
\section{Introduction}
\label{s.intro} \setcounter{equation}{0} \setcounter{footnote}{0}
%%%%%%%%%%%%%%%%%%%%%%%%%%%%%%%%%%%%%%%%%%%%%%%%%%%%%%%%%%%%%%%%%%%%%%
%%%%%%%%%%%%%%%%%%%%%%%%%%%%%%%%%%%%%%%%%%%%%%%%%%%%%%%%%%%%%%%%%%%%%
One of the most important unresolved questions in particle physics is how exactly the electroweak symmetry is broken. The standard model higgs mechanism provides ample motivation to come up with alternatives. An interesting new possibility is provided by higgsless models~\cite{gaugeinterval, realistichiggsless, sekhar, deconstruct, otherhiggsless} with a warped extra dimension~\cite{RandallSundrum}. A Higgs field localized on the IR brane of an RS background is decoupled by taking its VEV to be very large, while the masses of the $W$ and $Z$ bosons remain finite and are set by the size of the extra dimension. Unitarity of the gauge boson scattering amplitudes can then be ensured via heavy KK gauge boson exchange. Such models would solve the little hierarchy problem of Randall-Sundrum setups and have very distinctive phenomenological consequences. However, it is not clear whether these higgsless RS models can be made completely viable: a large correction to the S parameter makes it difficult to match electroweak precision data, the cutoff scale has to be adequately raised to ensure unitarization happens at weak coupling, and generically FCNC's are not adequately suppressed. Many of these initial difficulties have been at least partially addressed. One can tune the effective S-parameter away by making the fermion left-handed fermion wave functions close to flat~\cite{messytop}, and choosing the right fermion representations can prevent the large top mass from introducing coupling deviations in the $Zb\overline{b}$-vertex~\cite{custrep, custrephiggsless}. The cutoff scale can also be raised by lowering the curvature of the extra dimension \cite{messytop}. However, once the fermion wave functions are required to be close to flat, the traditional anarchic RS approach to flavor~\cite{RSflavmass ,RSflavCKM, RSGIM,  NMFV, DIU} (where fermion wave function overlaps generate fermion mass hierarchies and also give a protection called RS-GIM against FCNC's \cite{RSGIM}) can no longer be applied. A possible resolution to this problem is to introduce a genuine five-dimensional GIM mechanism, which uses bulk symmetries to suppress flavor violation \cite{5DGIM}. The trick is to impose global flavor symmetries on the bulk, with a large subgroup left unbroken on the IR brane and flavor mixing forbidden anywhere except the UV brane. One can then construct a model where tree-level FCNC's are genuinely vanishing, with the downside that we are no longer trying to explain the quark mass and mixing hierarchies, merely accommodating them.

The aim of this paper is to examine the flavor bounds (similar to~\cite{RSGIMreview, RSflavorbounds} on higgsless models and to present a viable flavor construction for these theories (see~\cite{RSotherflavor, RSleptons} for other examples in an RS context).  We have to circumvent the problems usually associated with higgsless models by ensuring that
\begin{tenumerate}
\item[\textbullet] all FCNCs are sufficiently suppressed,
\item[\textbullet] all tree-level electroweak precision constraints are satisfied,
\item[\textbullet] the cutoff scale is sufficiently high.
\end{tenumerate}
We show that the simplest versions of such a model cannot be realistic: imposing an exact GIM mechanism for all three generations either drives up the cutoff scale or prevents the S-parameter from being cancelled. Instead, the realistic flavor model we propose will have next-to minimal flavor violation (NMFV) \cite{5DGIM}, featuring a custodially protected quark representation for the third generation, and an exact GIM mechanism implemented for the first two generations only. This choice of representations allows us to isolate the lighter quarks from the dangerous top mass and prevent a large S-parameter without having to increase the bulk coupling and decrease the cutoff scale. Flavor-changing neutral currents are controlled by two main mechanisms:
\begin{tenumerate}
\item The surviving flavor symmetry between the first two generations forces all the mixing to go through the third generation (hence NMFV), which is vital to reduce $D$ and $K$ mixing.
\item Kinetic mixing terms on the UV confine the right-handed fermions to the UV brane and reduce bulk contributions to the couplings, which are the source of off-diagonal neutral couplings. This results in an RS-GIM-like flavor suppression mechanism for the right-handed fermions.
\end{tenumerate}
We also have some freedom to distribute the required charged-current mixing amongst the up- and down-sectors, which reduces neutral-current mixing in each sector. All of this is necessary to sufficiently suppress flavor violation. We find that experimental FCNC bounds systematically constrain the down-sector mixing angles, forcing them to lie within a volume of angle space that is enclosed by a well-defined surface. \emph{Assuming} the UV kinetic mixing terms obey a Cabibbo-type mixing hierarchy, this volume occupies $\sim O(5\%)$ of available angle space.

This paper is structured as follows: in \sref{background} we review the 5D GIM  mechanism and introduce the quark representations we will be using. In \sref{NMFV} we  outline our NMFV quark model and show compliance with electroweak precision data (EWPD). We also examine in detail the errors introduced by the zero mode approximation, and find that one can have zero S-parameter without flatness, provided there is a lot of KK mixing on the IR brane. The flavor suppression mechanisms of the NMFV model are derived in \sref{flavor} and demonstrated with the gluon KK contribution to FCNC's in \sref{FCNCbounds}. Numerical results for the mixing constraints are presented in \sref{numerical}, and we conclude with \sref{conclusion}.

%%%%%%%%%%%%%%%%%%%%%%%%%%%%%%%%%%%%%%%%%%%%%%%%%%%%%%%%%%%%%%%%%%%%%%
%%%%%%%%%%%%%%%%%%%%%%%%%%%%%%%%%%%%%%%%%%%%%%%%%%%%%%%%%%%%%%%%%%%%%%%
\section{Setup}
\label{s.background} \setcounter{equation}{0} \setcounter{footnote}{0}
%%%%%%%%%%%%%%%%%%%%%%%%%%%%%%%%%%%%%%%%%%%%%%%%%%%%%%%%%%%%%%%%%%%%%%
%%%%%%%%%%%%%%%%%%%%%%%%%%%%%%%%%%%%%%%%%%%%%%%%%%%%%%%%%%%%%%%%%%%%%%
After briefly reviewing the gauge sector, we will discuss the full 5D GIM mechanism \cite{5DGIM} and how one could apply it to various simple quark models. This will motivate the construction of our Next-to Minimal Flavor Violation Model in \sref{NMFV}.

\subsection{Gauge Sector}
\label{ss.gaugesector}
We work on an AdS$_5$ background and parameterize our space-time using conformal coordinates
\begin{equation}
ds^2=\left( \frac{R}{z}\right)^2 (dx_\mu dx_\nu \eta^{\mu\nu} -dz^2) \,
\end{equation}
where the UV and IR branes sit at $z = R, R'$ respectively. Our gauge sector takes the standard RS form \cite{firstflatness} with $\langle H \rangle \rightarrow \infty$ on the IR brane, as outlined in \cite{gaugesector}. This means we have an $SU(3)_c \times SU(2)_L \times SU(2)_R \times U(1)_X$ gauge group~\cite{firstflatness} in the bulk (where $X = B - L$), which is broken by boundary conditions to $SU(3)_c \times  SU(2)_L\times U(1)_Y$ on the UV brane and $SU(3)_c \times SU(2)_D \times U(1)_X$ on the IR brane. The custodial $SU(2)_R$ symmetry protects the $M_W/M_Z$ ratio from deviations at tree level, and the gauge boson mass is given by the size of the extra dimension. For future convenience, let us define
\begin{equation}
L \equiv \log R'/R.
\end{equation}
Then to leading order in $L$,
\begin{equation}
\label{e.Wmass}
M_W^2 \approx \frac{1}{R'^2 L}.
\end{equation}

On a technical note, we include all brane-localized gauge kinetic terms (BKTs) that are allowed by symmetries, and include their corrections due to 1-loop running effects on their respective branes \cite{gaugematching}. This means that the effective $SU(2)_{L}$ BKT on the UV brane can be negative at the weak scale. We will also localize some fermions on the UV brane, making the effective $U(1)_Y$ BKT always positive (but the effect is very small). Unless otherwise mentioned, we set all \emph{effective} BKTs to zero at the weak scale except for $U(1)_Y$, for which we set the bare term to zero. We also set the effective $SU(3)$ BKTs to zero, but note that they could be made negative at the weak scale.

In addition to the BKTs, the free input parameters in the gauge sector are $R$ and the ratio $g_{5R}/g_{5L}$. The SM $W, Z$ masses then determine $R'$ and $g_{5X}/g_{5L}$, while $\alpha_\mathrm{EM}$ and $\alpha_s$ at the weak scale set the overall size of the 5D couplings. All of our gauge bosons are canonically normalized, with all electroweak coupling corrections (including $S$ and $T$ parameters) pushed into interaction terms.

Our theory is valid up to a momentum cutoff, which in AdS$_5$ space is given conservatively by
\begin{equation}
\Lambda_\mathrm{cutoff} \sim \frac{16 \pi^2}{g_5^2} \frac{R}{R'}
\end{equation}
where $g_5$ is the largest 5-dimensional coupling. If we take $g_{5L} = g_{5R}$, then to leading order in $L$ the 4D coupling $g$ can be expressed as $g^2 = g_5^2/R L$ \cite{gaugesector}, which together with \eref{Wmass} gives
\begin{equation}
\label{e.cutoff}
\Lambda_\mathrm{cutoff} \sim \frac{16 \pi^2}{g^2} \frac{M_W}{\sqrt L} \approx \frac{29 \tev}{\sqrt L}.
\end{equation}
If we had a physical higgs on the IR brane we could freely choose our KK scale and make the cutoff large, but in the higgsless model we must choose a low curvature to unitarize $WW$-scattering before the theory becomes strongly coupled. We set $R^{-1} = 10^8 \gev$ which gives $L \approx 13$ and $\Lambda_\mathrm{cutoff} \sim 8 \tev$. We also choose $g_{5L} = g_{5R}$, since gauge matching would decrease the cutoff if we made the couplings different. The $Z'$ mass is therefore fixed and of order $\sim 700 \gev$.

\subsection{The Fermion Sector}
Our notation for a 5D Dirac fermion will be
\begin{equation}
\Psi = \left( \begin{array}{c} \chi \\ \overline{\psi} \end{array} \right)
\end{equation}
where $\chi$ and $\psi$ are both LH 2-component Weyl spinors, and hence $\overline\psi$ is a RH 2-component spinor. We write the boundary conditions for each Dirac spinor as $(\pm, \pm)$ at $z = (R, R')$, where  $+$ means $\overline \psi = 0$ and $-$ means $\chi = 0$. The normalized left-handed zero-mode profile is given by
\begin{equation}
\label{e.zeromode}
g_0(z) = R'^{-1/2} \left(\frac{z}{R}\right)^2 \left(\frac{z}{R'}\right)^{-c} f(c),  \ \ \ \ \  \mbox{where} \ \ \ \ \ f(c) = \sqrt \frac{{1-2c}}{1 - \left(\frac{R'}{R}\right)^{2c-1}}
\end{equation}
is the RS flavor function. The right-handed profile $f_0(z)$ is defined identically, with $c \rightarrow -c$.

\subsubsection*{5D GIM Mechanism}
For three generations of a given quark representation we can impose global flavor symmetries that prevent FCNCs at tree-level while generating all the required masses and mixings \cite{RSGIM}. In broad strokes, those symmetries must satisfy the following criteria:
\begin{tenumerate}
\item We need to have enough freedom to generate 6 different 4D quark masses and reproduce the CKM mixing matrix.
\item To ensure that flavor-violating operators are suppressed by the high scale $1/R$, we only allow flavor mixing on the UV brane, via right-handed kinetic terms for the up and down sector independently. Left-handed mixing is assumed to be forbidden by flavor symmetry, since otherwise the right- and left-handed kinetic terms cannot be diagonalized simultaneously and there would be FCNCs.
\item If we switch off the charged-current interactions we should be able to use the symmetries of the bulk and IR brane to diagonalize the UV mixing matrices. This makes the neutral currents diagonal in the 4D mass basis and forbids tree-level FCNCs.
\end{tenumerate}
We will now see how this mechanism can be applied to various simple quark models.

\subsubsection*{The Left-Right Symmetric Representation}
The simplest potentially realistic quark representation is the left-right symmetric representation, which has been adopted in~\cite{gaugesector, messytop, realistichiggsless}:
\begin{equation}
\label{e.QLQRdef}
Q_L = \left(\begin{array}{c} u \\ d\end{array} \right)_L  \sim  (2,1)_{1/6} \ \ \ \ \ \ \ \ \ \ Q_R = \left(\begin{array}{c} u \\ d\end{array} \right)_R  \sim  (1,2)_{1/6}
\end{equation}
We impose the boundary conditions $(+,+)$ on $u_L$  and $(-,-$) on $u_R$ to obtain left-handed and right-handed zero modes (similarly for $d$). On the IR brane, both $Q_L$ and $Q_R$ are $SU(2)_D$ doublets, so we can write a Dirac mass mixing term \cite{fermionbraneterms} to lift the zero modes,
\begin{equation}
\label{e.LRsymIRterm}
S_\mathrm{IR} = \int d^5x \left(\frac{R}{z}\right)^4 \delta(z - R') \ M_D  R' \ \overline{Q}_L Q_R + h.c.
\end{equation}
On the UV brane, the $Q_R$ doublet breaks up into two $SU(2)_L$ singlets, allowing us to assign brane-localized kinetic terms to $u_R$ and $d_R$ separately and supply the proper mass splitting.

If we want to implement the full 5D GIM mechanism using this representation, we populate the bulk with three copies of $Q_L, Q_R$ and impose flavor symmetries $G_\mathrm{bulk} = SU(3)_{Q_L} \times SU(3)_{Q_R}$ and $G_\mathrm{IR} = SU(3)_D$. This makes the bulk masses  $c_{Q_L}, c_{Q_R}$ and the IR Dirac mass $M_D$ flavor blind. On the UV brane we allow only kinetic mixing of the $Q_R$ fields, which take the form
\begin{equation}
\label{e.kineticterm}
\int d^5x \left(\frac{R}{z}\right)^4 \delta(z-R)   \psi^\alpha \sigma^\mu D_\mu K^{\alpha \beta} \overline{\psi}^\beta
\end{equation}
for both $u_R$ and $d_R$, where $K_u, K_d$ are two independent hermitian matrices and $\alpha, \beta$ are flavor indices. To forbid left-handed mixing, we impose the flavor symmetry $G_\mathrm{UV} = SU(3)_{Q_L} \times U(1)_{u_R} \times U(1)_{d_R}$.

If we switch off the charged-current interactions, the $u, d$ symmetries become independent and $G_\mathrm{IR} \rightarrow SU(3)_u \times SU(3)_d$, similarly in the bulk. These symmetries are broken on the UV brane, but we can use them to diagonalize the $K_u, K_d$ mixing matrices and end up with $G_\mathrm{UV} \supset U(1)_u^3 \times U(1)_d^3$, which prevents tree-level FCNCs. Another way to see this is as follows: since all other mass and kinetic terms in the action are flavor singlets, we can go to the 4D mass basis by rotating the fermions in flavor space and diagonalizing $K_{u,d}$. This pushes all the physical mixing into the charged-current couplings, giving us exactly the standard model CKM structure.

The main problem with this model is the top quark. To make it heavy, we must localize it close to the IR brane and make $M_D$ large. The flavor symmetry then forces \emph{all} the quarks to be close to the IR, generating a large negative $S$-parameter. The large flavor-blind $M_D$ has two additional dangerous effects: Firstly, the $L-R$ mixing causes left-handed quarks to live partially in the $R$ representation (and vice versa), which induces even more coupling corrections since it has the wrong quantum numbers. Secondly, the $KK$ mixing causes the light quarks to live partially in KK modes, resulting in dangerously high couplings to gauge KK modes.

One could try to address these problems by increasing the $SU(2)_D$ IR kinetic term, which corresponds to adding a positive bare $S$ parameter on the IR brane, but this is not viable. The matching of gauge couplings would force the coupling in the bulk to increase, lowering our cutoff to  $\sim O(1 \tev)$ and making the theory non-perturbative before the unitarization mechanism of higgsless RS models kicks in. We clearly need some way to protect the other quarks from deviations due to the heavy top mass.

\subsubsection*{The Custodially Protected Representation}
Focusing on the third generation only for a moment, the problem with the left-right symmetric representation is that the effects of the top mass are felt by the left-handed bottom (we can localize the right-handed bottom on the UV brane). Agashe et$.$ al~\cite{custrep} realized that deviations to those couplings could be avoided if
\begin{tenumerate}
\item[\textbullet] the $t_R$ is not in the same representation as any field which can mix with $b_L$, so that the top can have a separate IR Dirac mass that is not communicated to the bottom, and
\item[\textbullet] the representations that house the left-handed $b$ are symmetric under $SU(2)_L \leftrightarrow SU(2)_R$ interchange. This ensures that the $L$ and $R$ couplings are the same, meaning the $b_L$ couples to a linear combination of gauge bosons which is flat near the IR brane. Its couplings are therefore protected from deviations due to the $SU(2)_L \times SU(2)_R \rightarrow SU(2)_D$ breaking.
\end{tenumerate}
In the notation of~\cite{custrephiggsless}, the simplest representation which (almost) satisfies these requirements is:
\begin{equation}
\Psi_L = \left( \begin{array}{cc} X_L & t_L\\ T_L &b_L \end{array}\right)  \sim  (2,2)_{2/3} \ \ \ \ \ \
\Psi_R = \left(\begin{array}{ccc} X_R \\ T_R\\ b_R \end{array} \right) \sim (1,3)_{2/3} \ \ \ \ \ \
t_R  \sim  (1,1)_{2/3}
\end{equation}
For the $t,b$ quarks we impose the same boundary conditions as for the quarks in the left-right symmetric representation, while we make the exotic $T$, $X$ quarks (with electric charge $2/3$ and $5/3$ respectively) heavy by imposing mixed boundary conditions $(-,+)$ for $X_L, T_L$ and $(+,-)$ for $X_R, T_R$.

On the IR brane, the $\Psi_L$ bidoublet breaks down into an $SU(2)_D$ triplet and a singlet, which can mix with $\Psi_R$ and $t_R$.
\begin{equation}
\Psi_L^\mathrm{triplet} = \left( \begin{array}{c} X_L\\ \tilde{T}_L \\ b_L \end{array} \right) \equiv \left( \begin{array}{c} X_L\\ \sfrac{1}{\sqrt 2}(t_L + T_L) \\ b_L \end{array} \right), \ \ \ \ \ \ \ \ \ \ \Psi_L^\mathrm{singlet} = \tilde{t}_L \equiv \sfrac{1}{\sqrt 2} (t_L - T_L)
\end{equation}
Hence the allowed Dirac mass term on the IR brane is
\begin{equation}
\label{e.custodialIRterm}
S_\mathrm{IR} = \int d^5x \left(\frac{R}{z}\right)^4 \delta(z - R')\  R' \left[ M_3 \overline{\Psi}_R \Psi_L^\mathrm{triplet} + M_1 \overline{t}_R \Psi_L^\mathrm{singlet} + \mathrm{h.c.} \right],
\end{equation}
so we can make $M_1$ large to get a heavy top without influencing any field with the quantum numbers of the bottom. Also note that the custodial protection granted by this representation is not complete: KK-mixing via the $M_3$ mass term causes the left-handed $b$ to live partially in $b_R$, which is \emph{not} part of a $L\leftrightarrow{}R$ symmetric representation.\footnote{To make it $L\leftrightarrow{}R$ symmetric we would have to extend $\Psi_R$ to a $(1,3) \oplus (3,1)$.} This turns out to be good thing, since the fully protected coupling is a few percent too large in higgsless models (just the effect of the $S$-parameter). We can reduce it by localizing the $b_R$ closer to the UV brane, which increases $M_3$ (for a given 4D bottom mass) and hence increases KK-mixing. This in turn decreases the coupling, since it makes the LH bottom sensitive to the gauge boson profiles near the IR brane.

The unique features of this representation allow us to implement the full 5D GIM mechanism in a rather different fashion from the previous case. To protect the up and charm quarks from the heavy top, we make $M_1$ different for each quark generation and forbid all up-sector flavor mixing, including on the UV brane. The down sector symmetries, on the other hand, are chosen very similarly to the left-right symmetric representation: $M_3$ is flavor blind and large enough to generate the \emph{bottom} mass, and a kinetic mixing matrix $K_d$ on the UV brane generates quark mixing in the down sector (and hence all the physical quark mixing). This amounts to imposing the flavor symmetry  $G_\mathrm{bulk} = SU(3)_{\Psi_L} \times SU(3)_{\Psi_R} \times SU(3)_{u_R}$, which gets broken down to $G_\mathrm{IR} = SU(3)_\mathrm{triplets} \times U(1)_\mathrm{singlets}$ on the IR brane, and $G_\mathrm{UV} =  SU(3)_{\Psi_L} \times U(1)_{d_R} \times SU(3)_{u_R}$ on the UV brane (we also set all brane kinetic terms for $X, T$ fields to zero). FCNCs are prevented in the exact same fashion as for the left-right symmetric model, except we now only have to diagonalize the down sector.

Nevertheless, a higgsless quark model using \emph{only} the custodially protected representation is not viable. $\Psi_R$ must be close to the UV to match $g_{d_\ell^i d_\ell^i}$ to the SM. This fixes $c_{d_R}$ while leaving the $\Psi_L$ bulk mass $c_L$ unconstrained due to custodial protection. The problem is a tension between the LH up-type couplings and the RH bottom coupling. $g_{Z u_\ell^i u_\ell^i}$ can only be matched if $\Psi_L \rightarrow $ UV, since it is not protected and suffers large corrections near the IR brane. $g_{Z b_r b_r}$ has the opposite requirement. If $M_3$ is the minimal value to give $m_b$, $b_R$ lives entirely in the bulk. It must therefore be close to the UV brane to make it insensitive to the broken symmetry on the IR brane, which is not a problem. However, if $b_L \subset \Psi_L \rightarrow $ UV, then $M_3$ must be very large to generate $m_b$, which increases KK mixing and makes $g_{Z b_r b_r}$ sensitive to the IR brane again, reducing the coupling below SM. So while matching the LH up-type couplings to the SM requires $\Psi_L \rightarrow$ UV, the RH bottom coupling requires $\Psi_L \rightarrow $ IR. It is not possible to match both simultaneously. One might try to increase $M_3$ and confine the RH bottom to the UV brane, allowing $\Psi_L$ to be closer to the UV, but this also increases $T$-mixing with the up-type quarks, which forces $\Psi_L$ \emph{even closer} to the UV to get a match. One cannot achieve overlap.

So while this model can protect the left-handed bottom couplings, the 5D GIM mechanism forces \emph{all} the up-type quarks to behave like the troublesome top, and their couplings cannot be matched to the SM simultaneously with the RH bottom.

%%%%%%%%%%%%%%%%%%%%%%%%%%%%%%%%%%%%%%%%%%%%%%%%%%%%%%%%%%%%%%%%%%%%%%
%%%%%%%%%%%%%%%%%%%%%%%%%%%%%%%%%%%%%%%%%%%%%%%%%%%%%%%%%%%%%%%%%%%%%%%
\section{The NMFV Quark Model}
\label{s.NMFV} \setcounter{equation}{0} \setcounter{footnote}{0}
%%%%%%%%%%%%%%%%%%%%%%%%%%%%%%%%%%%%%%%%%%%%%%%%%%%%%%%%%%%%%%%%%%%%%%
%%%%%%%%%%%%%%%%%%%%%%%%%%%%%%%%%%%%%%%%%%%%%%%%%%%%%%%%%%%%%%%%%%%%%%

The complete 5D GIM mechanism is too restrictive for higgsless RS model-building. We have to give up a some flavor protection in exchange for agreement with electroweak precision data, while ensuring that FCNCs are still under control. This motivates us to combine both representations in a single quark model with \emph{next-to minimal flavor violation}, harnessing their complementing strengths while keeping as much flavor symmetry as possible.

\subsection{Setup}
\label{ss.nmfvsetup}
Two copies of $Q_L, Q_R$ with bulk masses $c_{Q_L}, c_{Q_R}$ make up the first two generations, while the third generation is contained in the custodially protected $\Psi_L, \Psi_R, t_R$ with bulk masses $c_L, c_{b_R}, c_{t_R}$. This protects the other quarks from the influence of the heavy top while enabling us to match all fermion couplings to experimental data. (Note that the top couplings are poorly constrained.) The form of the respective IR Dirac mass terms are given in equations (\ref{e.LRsymIRterm}) and (\ref{e.custodialIRterm}). We impose the flavor symmetry $G_\mathrm{bulk} = SU(2)_{Q_L} \times SU(2)_{Q_R}$ in the bulk, which is broken down to $G_\mathrm{IR} = SU(2)_D$ on the IR brane. This means that the first two generations have the same IR Dirac mass $M_D$ and bulk masses. The third generation has the IR Dirac masses $M_3$ for the $SU(2)_D$ triplet (which includes the bottom) and $M_1$ for the singlet (which supplies mass to the top). To provide flavor mixing and differentiate the quark masses of the first two generations, we must introduce general hermitian $3 \times 3$ kinetic mixing matrices $K_u$ and $K_d$ as in \eref{kineticterm}. Therefore, the flavor symmetry on the UV brane is $G_\mathrm{UV} = SU(3)_{Q_L} \times U(1)_{u_R} \times U(1)_{d_R}$ (where the third $Q_L$ is contained in $\Psi_L$).

We can see immediately that there will be FCNCs in this model. The flavor symmetry is explicitly broken by choosing a different quark representation for the third generation. If we switch off the charged currents, we only have $SU(2)$ symmetries available, which are not enough to diagonalize the kinetic mixing matrices on the UV brane. However, as we will see, this partial symmetry is enough to force all mixing to go `through the third generation' and suppress 12-mixings.

\subsection{Going to 4D Mass Basis}
\label{ss.4Dmassbasis}
We can solve the bulk equations with the appropriate BC's to compute the entire KK tower of fermion wave functions. After integrating out the 5$^\mathrm{th}$ dimension, we end up with a 4D action containing the following terms (using matrix notation in flavor/KK space):
\begin{align}
\label{e.matrixnotations}
\begin{array}{rl}
\mbox{4D mass terms}  & \ \  \psi_{u,d} M_{u,d} \chi_{u,d},\\
\mbox{RH kinetic mixing terms} & \ \ \psi_{u,d} \ \sigma^\mu (\mathbbm{1} + f_{u,d} K_{u,d} f_{u,d}) \partial_\mu \overline{\psi}_{u,d} \equiv \psi_{u,d} \ \sigma^\mu \kappa_{u,d} \partial_\mu \overline{\psi}_{u,d},\\
\mbox{coupling terms like} & \ \ \overline{\chi}_u \overline{\sigma}^\mu Z_\mu^{(n)} g_{Z u_L u_L}^{(n)} \chi_u,
\end{array}
\end{align}
where $(n)$ is a gauge boson KK index, $f_{u,d}$ is a diagonal matrix of the right-handed fermion wave functions evaluated at $z = R$, and $K_{u,d}$ is the UV brane kinetic mixing matrix.

To go to 4D mass basis, we must first diagonalize and canonically normalize the kinetic mixing term by rotating the RH spinors with a hermitian matrix $H$. Once the kinetic terms are flavor singlets we can diagonalize the mass matrices with the usual biunitary transformation. We will always distinguish quantities in the physical basis with a `mass' superscript from quantities in the original flavor basis without superscript. The quark spinors in the mass basis are related to the flavor basis in the following way:
\begin{equation}
\label{e.spinorbasischange}
\begin{array}{rlcrl}
\chi_u &= U_{Lu} \chi_u^\mathrm{mass} && \chi_d &= U_{Ld} \chi_u^\mathrm{mass}\\
\overline{\psi}_u &= H_u U_{Ru} \overline{\psi}_u^\mathrm{mass}  & & \overline{\psi}_d &= H_d U_{Rd} \overline{\psi}^\mathrm{mass}_d.
\end{array}
\end{equation}
Applying this transformation to the the mass terms, the left/right-handed neutral couplings (denoted generically by $g_L$/$g_R$), and the left/right-handed $W$ couplings, we get:
\begin{equation}
\label{e.basischange}
\begin{array}{rlcrl}
M_u^\mathrm{mass} &= U_{Ru}^\dagger H_u^\dagger M_u U_{Lu} && M_d^\mathrm{mass} &= U_{Rd}^\dagger H_d^\dagger M_d U_{Ld}\\
g_{Lu}^\mathrm{mass} &= U_{Lu}^\dagger g_{Lu}U_{Lu} && g_{Ru}^\mathrm{mass} &= U_{Ru}^\dagger H_u^\dagger g_{Ru}H_u U_{Ru} \\
g_{Ld}^\mathrm{mass} &= U_{Ld}^\dagger g_{Ld}U_{Ld}&& g_{Rd}^\mathrm{mass} &= U_{Rd}^\dagger H_d^\dagger g_{Rd}H_d U_{Rd}\\
g_{W u_L d_L}^\mathrm{mass} &= U_{Lu}^\dagger g_{W u_L d_L} U_{Ld} &&
g_{W u_R d_R}^\mathrm{mass} &= U_{Ru}^\dagger H_u^\dagger g_{W u_R d_R}H_d U_{Rd}
\end{array}
\end{equation}
There is a very useful relation which we will need later. We simply write out $|M^\mathrm{mass}|^2 = {M^\mathrm{mass}}^\dagger M^\mathrm{mass} = M^\mathrm{mass} {M^\mathrm{mass}}^\dagger$ (since $M^\mathrm{mass}$ is diagonal). Keeping in mind that the $H$ matrices are hermitian and $H^2 = \kappa^{-1}$, we find
\begin{align}
\label{e.getkappa}
\left| M_u^\mathrm{mass} \right|^2 &= U_{Lu}^\dagger \left( M_u^\dagger \kappa_u^{-1} M_u \right) U_{Lu} =U_{Ru}^\dagger H_u^\dagger M_u M_u^\dagger H_u  U_{Ru}\\
\nonumber \left| M_d^\mathrm{mass} \right|^2 &= U_{Ld}^\dagger \left( M_d^\dagger \kappa_d^{-1} M_d \right) U_{Ld} =U_{Rd}^\dagger H_d^\dagger M_d M_d^\dagger H_d  U_{Rd}.
\end{align}

The exotic X-quark with charge $5/3$ is an interesting experimental signature of our model. Its mass is roughly half a $\tev$ and it couples to the top  via charged-current interactions (in the flavor basis) with coupling strength comparable to but generically less than $g/\sqrt 2$. The coupling in the mass basis is
\begin{equation}
\label{e.Xcoupling}
\begin{array}{rl}
g_{W X_L u_L}^\mathrm{mass} &= g_{W X_L u_L} U_{Lu} \\
g_{W X_R u_R}^\mathrm{mass} &=  g_{W X_R u_R}H_u U_{Ru}.
\end{array}
\end{equation}
Detection could be possible at the LHC with less than $100$ pb$^{-1}$ of integrated luminosity \cite{Xquark}.

\subsection{Satisfying Electroweak Precision Data and CDF Bounds}
\label{ss.CDF}
It is not hard to see why this model can satisfy electroweak precision constraints. The heavy top mass does not influence the other quarks, and the correct bottom couplings can be achieved by moving $\Psi_L \supset t_L, b_L $ and $t_R$ close to the IR brane, while the $\Psi_R \supset b_R$ is close to the UV \cite{custrephiggsless}. The top couplings will deviate from the SM value, but this is acceptable since it is poorly constrained experimentally. The first- and second-generation couplings can be made to agree with the SM by adjusting \ $c_{Q_L}, c_{Q_R} \sim 0.5$, and we have enough freedom to choose IR Dirac masses and UV kinetic terms to generate all the different quark masses and mixings. It is worth noting that the $Q_{L,R}$ bulk masses can take on a range of values, due to the effect of KK-mixing which we will discuss in \ssref{zeromode}. We explicitly demonstrated EWPD compliance using two different numerical calculations. In the first, we assumed that there is no flavor mixing and absorbed the diagonal boundary terms into BC's. In the second there was flavor mixing, and we followed the procedure of \ssref{4Dmassbasis}: using the zero mode approximation in which the boundary terms act as mixing terms between zero modes and KK modes.
%%%%%%%%%%%%%%%%%%%%%%%%%%%%%

One of the canonical signatures of higgsless models are light gauge KK-modes with a mass of $\approx 700 \gev$. This is low enough to warrant closer inspection of current CDF bounds \cite{CDF,CDFttZ, CDFttG} to make sure our model is not already excluded. The CDF searches for heavy gauge bosons focus on resonant pair production processes of the form (light quark pair) $\rightarrow$ (heavy gauge boson) $\rightarrow$ (some fermion pair, e.g. $e \overline e, t \overline b$). Assuming that the coupling to the heavy gauge boson is the same as to the SM counterpart for both the initial and final fermion states, the CDF bounds are $m_{W'}, m_{Z'} \gsim 800 \gev$. However, those bounds must be adjusted for our model since the coupling of gauge KK modes is very suppressed for the first two quark generations, and somewhat enhanced for the third generation.
\vspace{-0.1cm} \begin{equation}
\begin{array}{|c|c|c|cc|}
\hline
\mbox{quark generation}  & \multicolumn{4}{l|}{\mbox{approx. coupling as a multiple of SM}}\\
& Z' & W' & G' & \\
\hline
1,2 \mbox{ (LH)} & < 1/5  & 1/100 & < 1/4 & \\
1,2 \mbox{ (RH)} & 1/5 & 1/100 & 1/4 &\\
3 & 2-4 & 1 & 2& \\
\hline
\end{array}
\end{equation}
Since the light left-handed quarks are not UV-localized, their couplings depend sensitively on the bulk masses and can be very small. Leptons in our model would have similar couplings to the light quarks. It is clear that the coupling suppression increases the $m_{W'}, m_{Z'}$ bounds from leptonic and $tb$-channel searches way beyond our KK-scale of 700 GeV. Due to low $t \overline{t}$-detection efficiencies, the $t\overline{t}$-channel also does not supply a meaningful $m_{Z'}$ bound \cite{CDFttZ}.

Only the constraints on $m_{G'}$ from \cite{CDFttG} require closer inspection. Their analysis assumed vector-like couplings to $G'$ which were parameterized as $g_{\mathrm{light\, quarks}} = \lambda_q g_s$ and $g_{top} = \lambda_Q g_s$. The bounds on $m_{G'}$ depend on $\lambda = \lambda_g \lambda_Q$ and the width $\Gamma$ of $G'$. If we assume that we can use those bounds for our \emph{chiral} couplings by simply averaging and setting $\lambda_q = \sfrac{1}{2}(\lambda_{q_L} + \lambda_{q_R}) \approx 0.25 - 0.5$, we can extract an approximate bound of $\Gamma/m_{G'} \gsim 0.2$ on the width of our KK-gluon if its mass is $ \approx 700 \gev$. We have not calculated the width of the $G'$ since it depends on several parameters that are not completely fixed in our model, but $\Gamma/m_{G'} \sim 0.2$ is not an atypical value for RS KK-gluons, see for example \cite{RandallKKGluon}. Furthermore, we can also decrease $\lambda$ by another factor of $\sim 4$ by taking into account 1-loop RGE corrections to the $SU(3)_c$ UV brane kinetic term, as outlined in \ssref{gaugesector}. This alleviates any concern that our model might be excluded by CDF bounds. However, the relatively light $G'$ should certainly be detected at the LHC.

\subsection{Counting Physical Parameters and the Meaning of Large UV Kinetic Terms}
Each $N \times N$ hermitian UV kinetic mixing matrix $K_u, K_d$ is defined by $N^2$ parameters, $N(N+1)$ real elements and $N(N-1)$ complex phases. For $N = 3$, this gives a total of 12 real parameters and 6 phases. We can always do an $SU(2)\times U(1)$ flavor rotation, which corresponds to eliminating unphysical parameters: it removes 1 angle and 3 phases. This leaves us with 11 real parameters and 3 phases, which includes the 6 quark masses. Hence the parameters in the flavor sector are 6 quark masses, 5 mixing angles and 3 phases, as well as 3 IR Dirac masses $M_D, M_1$ and $M_3$.

At this point a remark about the \emph{size} of the UV kinetic terms is in order. The $K_{u,d}$ matrix elements will be very large, generically $\sim [O(10^2) - O(10^9)] R$, but this is no cause for concern. After canonically normalizing, the magnitude of the $K$'s will merely specify what fraction of the fermions lives on the UV brane (i.e. is elementary in the AdS/CFT picture), and how much lives in the bulk (i.e. is composite). In our model the right-handed quarks are almost entirely confined to the UV brane, only slightly dipping into the bulk to mix with the left-handed quarks on the IR brane and generate a Dirac mass.

\subsection{The Zero Mode Calculation}
\label{ss.zeromode}
The fermion zero mode approximation enormously simplifies matching and mixing calculations, and we can use it to gain a great deal of insight into the flavor protection mechanisms of our model. However, KK mixing is much more significant for higgsless models than for standard RS with multi-TeV KK masses, so we need to investigate the range of validity of this approximation in detail if we want to trust our calculations.

\subsubsection*{Error Estimate}
Consider a simple toy-model with a single generation of quarks in the left-right symmetric representation \eref{QLQRdef}. There is a Dirac mass term on the IR brane (\ref{e.LRsymIRterm}) and a UV boundary kinetic term for the right-handed fields (\ref{e.kineticterm}). Focusing only on the \emph{left-handed} fields for the moment, we can incorporate the IR Dirac mass term into the $z = R'$ boundary conditions of the 5D wave function profiles \cite{fermionbraneterms}, eliminating KK-mixing on the IR brane:
\begin{equation}
\label{e.IRBC}
g_{u_R} = R' M_D g_{u_L} |_{z = R'},
\end{equation}
similarly for the down sector. We will assume that the errors are small and $c_{Q_L}, -c_{Q_R} > 0$.

If there was no IR mixing, $g_{u_L}$ would just be the zero mode $g_{c_{Q_L}}^0$ (i.e. $g_0$ from \eref{zeromode} with $c \rightarrow c_{Q_L}$), and adding a small amount of mixing should not change the shape of that waveform significantly. The new mode that appears due to mixing is ${g_{u_R}}$, and its shape is also independent of the size of a small mass. Hence it should be the zero mode that is normally projected out when the BC's do not include any mixing, i.e. $g_{c_{Q_R}}^0$ (which is different from the usual RH zero mode $f_{c_{Q_R}}^0 = g_{-c_{Q_R}}^0$). A simple ansatz to approximately solve the exact BCs is therefore
\begin{equation}
\label{e.gLgRansatz}
g_{u_L} = a g_{c_{Q_L}}^0 \ \ \ g_{u_R} = b g_{c_{Q_R}}^0.
\end{equation}
Using the fermion normalization condition $\int dz (R/z)^4 (|g_{u_R}|^2 + |g_{u_L}|^2) = 1$ as well as \eref{IRBC}, we can solve for the coefficients $a$ and $b$. Assuming the error is small, one obtains
\begin{equation}
\label{e.absoln}
a \approx 1 - \frac{1}{2}\left( R' M_D
\frac{f(c_{Q_L})}{f(c_{Q_R})}\right)^2 \  \ \ \ \ \ b \approx \left( R' M_D
\frac{f(c_{Q_L})}{f(c_{Q_R})}\right)
\end{equation}
We can now estimate the deviation of a typical coupling to gauge boson $\Psi$ compared to the zero mode approximation:
\begin{equation}
\int dz \left( \frac{R}{z} \right)^4 |g_{u_L}|^2 g_5 \Psi
= \left[1 - \left( R' M_D \frac{f(c_{Q_L})}{f(c_{Q_R})}\right)^2\right] \int dz \left( \frac{R}{z} \right)^4 |g^0_{u_L}|^2 g_5 \Psi
\end{equation}
The correction due to including the $g_R$ is at most of similar order, and in fact much smaller for electroweak couplings since $|\Psi^{R3}| < |\Psi^{L3}|$ near the IR brane and $g_{c_{Q_R}}^0$ is extremely IR localized. Hence the zero mode approximation \emph{overestimates} left-handed couplings by roughly
\begin{equation}
\label{e.deltaL}
\delta_L \sim \left( R' M_D \frac{f(c_{Q_L})}{f(c_{Q_R})}\right)^2,
\end{equation}
which is a \emph{relative} error independent of the gauge charge. By a similar procedure we obtain the error for the right-handed couplings. It is simplest to \emph{not} include the UV brane term in the BCs and simply renormalize the bulk wave function. Thus we find that the zero mode approximation \emph{overestimates} right-handed couplings by
\begin{equation}
\label{e.deltaR}
\delta_R \sim \left(\frac{ R' M_D}{\sqrt{1 + K {f^0_{u_R}(R)}^2}} \frac{f(-c_{Q_R})}{f(-c_{Q_L})}\right)^2.
\end{equation}
which is negligible unless the UV term is very small. Both of these error estimates have been confirmed numerically. Using \eref{Wmass} we can express them as
\begin{equation}
\label{e.deltaLR}
\delta_L \sim \frac{M_D^2}{M_W^2 \log R'/R} \frac{f(c_{Q_L})^2}{f(c_{Q_R})^2}
\ \ \ \ \ \ \ \delta_R \sim \frac{1}{1 + K {f^0_{u_R}(R)}^2} \frac{M_D^2}{M_W^2 \log R'/R} \frac{f(-c_{Q_R})^2}{f(-c_{Q_L})^2}.
\end{equation}

\begin{figure}
\begin{center}
\includegraphics[keepaspectratio, width = 15cm]{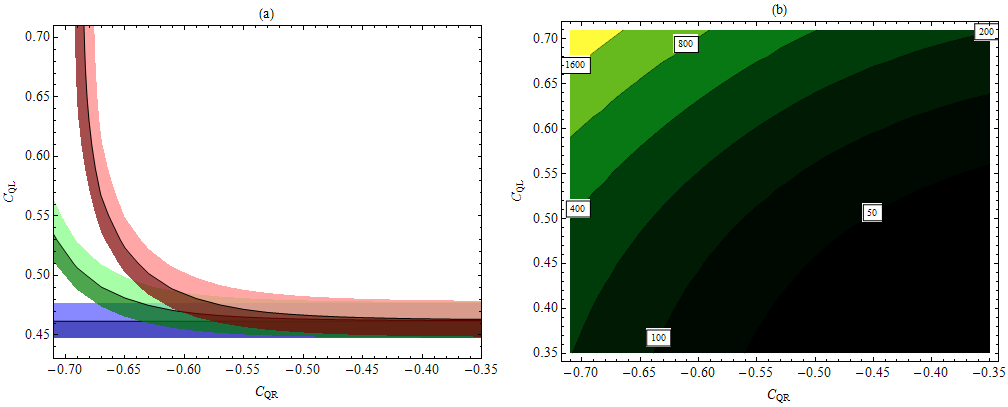}
\caption{(a) shows the region of bulk masses that reproduces the SM couplings for the first quark generation with varying magnitudes of the IR brane mass. The three bands are for three values of $\rho_d = M_D/M_D^\mathrm{min}$, where $M_D^\mathrm{min}$ is the smallest possible IR brane Dirac mass which can reproduce the first generation masses. Darker (lighter) regions indicate that the coupling is up to 0.6\% below (above) SM value. The blue band for $\rho_d = 1$ is reproduced by the zero mode approximation, and shows that $Q_L$ must be almost flat and near the IR brane as expected. The green and red bands correspond to $\rho_d = 600$ and $1000$, and we see significant shifts which allow the quarks to be UV localized. (b) shows $M_D^\mathrm{min}$ for the first generation in MeV. It is clear that increasing those brane terms by a factor of 1000 is not necessarily unreasonable, since it corresponds to a TeV scale $M_D$.}\label{f.1stgenSMmatchdiffros}
\end{center}
\end{figure}
To demonstrate how significant those errors can be, we computed the couplings in our toy model numerically for the first quark generation only, incorporating \emph{both} UV and IR brane terms into boundary conditions. As \fref{1stgenSMmatchdiffros} shows, we find that one can now have \emph{both} $Q_L$ and $Q_R$ localized near the UV brane without any $S$-parameter by turning up the value of $M_D$! Thus one can get around the canonical assumption that fermions in higgsless RS models must be almost flat and near the IR brane. If there is significant KK mixing on the IR brane, the fermions can be UV localized.

\subsubsection*{Using the Zero Mode Approximation for our Model}
As we discussed in \ssref{CDF}, we have used more accurate calculational methods to confirm that our model can satisfy electroweak precision constraints. The zero mode calculation without fermion KK-modes will only be used to estimate FCNC suppression. We must expect percent-level errors for all diagonal couplings and masses that do not involve the top, and $O(1)$ errors for the top mass and off-diagonal couplings in the up sector. Going from the zero mode calculation to full KK mixing preserves the general mixing hierarchy, but we would have to restore a full CKM match by adjusting the up-sector mixing angles by order unity. This level of accuracy is sufficient to estimate the tightly constrained down-sector FCNCs to within a few percent. Our estimate for D-mixing, on the other hand, will only be valid up to a factor of order unity, but this is enough to demonstrate our flavor suppression mechanism.

%%%%%%%%%%%%%%%%%%%%%%%%%%%%%%%%%%%%%%%%%%%%%%%%%%%%%%%%%%%%%%%%%%%%%%
%%%%%%%%%%%%%%%%%%%%%%%%%%%%%%%%%%%%%%%%%%%%%%%%%%%%%%%%%%%%%%%%%%%%%%%
\section{Flavor Matching and Protection in the NMFV Model}
\label{s.flavor} \setcounter{equation}{0} \setcounter{footnote}{0}
%%%%%%%%%%%%%%%%%%%%%%%%%%%%%%%%%%%%%%%%%%%%%%%%%%%%%%%%%%%%%%%%%%%%%%
%%%%%%%%%%%%%%%%%%%%%%%%%%%%%%%%%%%%%%%%%%%%%%%%%%%%%%%%%%%%%%%%%%%%%%
We will now analyze the flavor-protection mechanisms of the NMFV model in the framework of the zero-mode calculation. Our first task is to find the correct UV-localized right-handed kinetic mixing matrices $K_u$ and $K_d$ which reproduce the 4D CKM matrix. After obtaining a tree-level match to the Standard Model we proceed to find the off-diagonal neutral couplings which give rise to dangerous tree-level FCNCs.

\subsection{Matching in the Zero Mode Approximation}
Throughout this section we will drop the $u$, $d$ subscripts whenever the derivations for the up- and down-sectors are identical. The problem of matching the zero mode calculation to the Standard Model factorizes into four steps:
\begin{tenumerate}
\item Find bulk masses $c_{Q_R}, c_{Q_L}, c_{b_R}, c_{t_R}$ and $c_L$ which give the correct quark couplings (noting that some deviation from SM is permissable for the top).
\item Choose a set of unitary matrices $U_{Lu}$, $U_{Ld}$ that match $g_{W u_L d_L}^\mathrm{mass} = U_{Lu}^\dagger g_{W u_L d_L} U_{Ld}$ to the experimental value of $\sfrac{g}{\sqrt 2} V_\mathrm{CKM}$. There is a lot of freedom to choose mixing matrices here, and we shall address it in \sref{numerical}.
\item Choose IR Dirac masses $M_D, M_3$ and $M_1$ which are at least big enough to supply the charm, bottom and top masses, and bigger if we want to confine the RH quarks to the UV brane.
\item Find the $K_u, K_d$ kinetic matrices which are required to produce the SM masses and mixings.
\end{tenumerate}
The fourth step works as follows. The total kinetic term is $\kappa = \mathbbm{1} + f_R K f_R$ (see eqn. \ref{e.matrixnotations}), where $f_R = \mathrm{Diag}(f_0^1, f_0^2, f_0^3)|_{z = R'}$ is the diagonal matrix of RH quark zero modes evaluated on the UV brane (note that $f_0^1 = f_0^2$ due to flavor symmetry). Using \eref{getkappa} we can express $\kappa$ in terms of quantities that we know:
\begin{equation}
\left| M^\mathrm{mass} \right|^2 = U_L^\dagger (M^\dagger \kappa^{-1} M) U_L \phantom{spacer}
\Longrightarrow
\phantom{spacer} \kappa =  M U_L \left| M^\mathrm{mass} \right|^{-2} U_L^\dagger M^\dagger.
\end{equation}
Hence we obtain an expression for the UV brane kinetic mixing matrix:
\begin{equation}
\label{e.findK}
K = f_R^{-1} \left[ M U_L \left| M^\mathrm{mass} \right|^{-2} U_L^\dagger M^\dagger - 1 \right] f_R^{-1}
\end{equation}

\subsection{Flavor Protection}
In the flavor basis, we can always write any left-handed neutral coupling as:
\begin{equation}
\label{e.gLflavorbasis}
g_L = (g \Psi)_L \cdot \mathbbm{1} + g^\mathrm{bulk}_L,
\end{equation}
where $(g \Psi)_L$ is the wave function of the gauge boson that the fermion couples to, evaluated on the UV brane and multiplied by the appropriate gauge coupling (in the AdS/CFT picture, this is the elementary part of the gauge boson). All the flavor non-universalities are contained in the diagonal matrix $g^\mathrm{bulk}_L$ which comes from bulk overlap integrals with the fermions (and corresponds to the composite gauge boson coupling). For a right-handed neutral coupling we also have the contribution from the UV kinetic term, which gives
\begin{equation}
\label{e.gRflavorbasis}
g_R \ = \ (g \Psi)_R \cdot (\mathbbm{1} + f_R K f_R)  + g^\mathrm{bulk}_R \ =  \ (g \Psi)_R \cdot \kappa + g^\mathrm{bulk}_R.
\end{equation}
The form of $K$ is known from \eref{findK}, and we can obtain the right-handed rotation matrix from \eref{basischange}
\begin{equation}
\label{e.HUR}
H U_R = \left( M^\mathrm{mass} U_L^\dagger M^{-1} \right)^\dagger.
\end{equation}
The left-handed couplings just rotate by $U_L$. This is all the information we need to transform the couplings  into the physical basis, where the mass matrix is $M^\mathrm{mass} = \mathrm{Diag}(m_1^{\mathrm{SM}},m_2^{\mathrm{SM}},m_3^{\mathrm{SM}})$.

Before we do that, however, it is useful to parameterize the IR Dirac masses in terms of roughly how large we want the UV kinetic terms to be, i.e. how strongly we want to confine the RH quarks to the UV brane. In the flavor basis, the mass matrix is
\begin{equation}
M = \mathrm{Diag}(M_1, M_1, M_3),
\end{equation}
where the flavor symmetry forces the first two terms to be the same and
\begin{equation}
\begin{array}{rclcl}
M_1^{u,d} & = & f(-c_{Q_R}) f(c_{Q_L}) M_D & \equiv &\rho_c m_c \\
M_3^d & = &  f(-c_{b_R}) f(c_L) M_3 & \equiv &\rho_b m_b \\
M_3^u  & = &  f(-c_{t_R}) f(c_L) M_1/ \sqrt 2 & \equiv & \rho_t m_t.
\end{array}
\end{equation}
$\rho_{c,b,t}  = 1$ corresponds to choosing the minimal IR Dirac mass (and a correspondingly minimal UV kinetic term) which can generate the $c,b,t$ 4D mass. $\rho_{c,b,t} > 1$ simply corresponds to increasing the IR Dirac mass by that factor, which  also increases the UV kinetic term in order to keep the quark mass constant. This localizes the RH quark on the UV brane.

Now we can apply the basis transformations $U_L$ and $H U_R$ to eqns. (\ref{e.gLflavorbasis}) and (\ref{e.gRflavorbasis}). We obtain expressions for the physical 4D neutral couplings:
\begin{align}
\label{e.gLgR}
\nonumber g_R^\mathrm{mass} &= (g \Psi)_R \cdot \mathbbm{1} + M^\mathrm{mass} U_L^\dagger M^{-1} g_R^\mathrm{bulk} (M^\dagger)^{-1} U_L {M^\mathrm{mass}}^\dagger\\
g_L^\mathrm{mass} &= (g \Psi)_L \cdot \mathbbm{1} +  U_L^\dagger g_L^\mathrm{bulk} U_L
\end{align}
All the off-diagonal terms come from the flavor non-universal bulk part of the coupling, rotated by the appropriate transformation matrix.

Let us now find explicit expressions for these off-diagonal neutral coupling elements. Our flavor symmetry imposes $g_L^\mathrm{bulk} \equiv \mathrm{Diag}(g_L^{\mathrm{bulk}1}, g_L^{\mathrm{bulk}1}, g_L^{\mathrm{bulk}3})$ and $g_R^\mathrm{bulk} \equiv \mathrm{Diag}(g_R^{\mathrm{bulk}1}, g_R^{\mathrm{bulk}1}, g_R^{\mathrm{bulk}3})$. The most general form (ignoring phases) that $U_L$ can take is:
\begin{equation}
\label{e.ULnophases}
U_L =
\left(
\begin{array}{ccc}
 c_{12} c_{13} & c_{13} s_{12} & s_{13} \\
 -c_{23} s_{12}-c_{12} s_{13} s_{23} & c_{12} c_{23}-s_{12} s_{13} s_{23} & c_{13} s_{23} \\
 -c_{12} c_{23} s_{13}+s_{12} s_{23} & -c_{23} s_{12} s_{13}-c_{12} s_{23} & c_{13} c_{23}
\end{array}
\right)
\end{equation}
If we assume completely anarchic UV mixing, the FCNC's will generically be too large. However, if one assumes Cabibbo-type hierarchies for both $U_{Lu}$ and $U_{Ld}$ mixing matrices,
\begin{eqnarray}
s_{12} = O(1) \times \lambda \ \ \ \ s_{23} = O(1) \times \lambda^2 \ \ \ \ s_{13} = O(1) \times \lambda^3  \ \ \ \ \lambda \sim 0.2,
\end{eqnarray}
then the flavor-changing effects will get an additional Cabibbo-suppression. This amounts to assuming that there is some systematic UV physics generating the mixing hierarchies. Substituting all this into \eref{gLgR} and expanding to lowest order in $\lambda$, we obtain simple expressions for the off-diagonal neutral couplings. In the down sector,
\begin{equation}
\label{e.offdiagNC}
\begin{array}{ccc}
\begin{array}{rcl}
{g_{Ld,ij}^\mathrm{mass}} &\approx &  \left[g_{Ld}^{\mathrm{bulk}3} - g_{Ld}^{\mathrm{bulk}1}\right] U_{Ld}^{ji} \\ \\
{g_{Rd,ij}^\mathrm{mass}} &\approx & \left[\dfrac{g_{Rd}^{\mathrm{bulk}3}}{\rho_{b}^2 m_{b}^2} - \dfrac{g_{Rd}^{\mathrm{bulk}1}}{\rho_c^2 m_c^2}\right]  m_d^j m_d^i U_{Ld}^{ji}
\medskip
\end{array}
&
\phantom{spacer}
&
\begin{array}{l}
\mbox{where } i < j\\ \\
\mbox{and for } (i,j) = (1,2), \medskip \\ U_{Ld}^{ji} \rightarrow U_{Ld}^{32} U_{Ld}^{31}.
\end{array}
\end{array}
\end{equation}
For the up sector just change subscripts $d \rightarrow u$ and replace $\rho_b m_b$ by $\rho_t m_t$.

The flavor protection of our model is now apparent. Firstly, the surviving \emph{flavor symmetry} between the first two generations forces all the mixing to go through the third generation (hence NMFV). This is vital to push $D$ and $K$ mixing below the stringent experimental bounds. Secondly, since we are free to increase $\rho > 1$, there is an \emph{RS-GIM-like flavor suppression mechanism} for the right-handed fermion couplings. This is due to the kinetic mixing terms, which confine the right-handed quarks to the UV brane and suppress the bulk contributions to the couplings, which are the source of flavor violation. Finally, since the charged-current mixing matrix is made up of both the up- and down-sector mixing matrices, we have some freedom to `divide up the mixing'  between the two sectors and reduce FCNCs for each sector accordingly.

%%%%%%%%%%%%%%%%%%%%%%%%%%%%%%%%%%%%%%%%%%%%%%%%%%%%%%%%%%%%%%%%%%%%%%
%%%%%%%%%%%%%%%%%%%%%%%%%%%%%%%%%%%%%%%%%%%%%%%%%%%%%%%%%%%%%%%%%%%%%%%
\section{Estimating FCNCs for the NMFV Model}
\label{s.FCNCbounds} \setcounter{equation}{0} \setcounter{footnote}{0}
%%%%%%%%%%%%%%%%%%%%%%%%%%%%%%%%%%%%%%%%%%%%%%%%%%%%%%%%%%%%%%%%%%%%%%
%%%%%%%%%%%%%%%%%%%%%%%%%%%%%%%%%%%%%%%%%%%%%%%%%%%%%%%%%%%%%%%%%%%%%%

We are now in a position to estimate the FCNCs for our NMFV model and compare them to experimental bounds, as well as to a standard RS setup with a KK scale of $3 \tev$ and anarchic Yukawa couplings, where the only flavor protection is due to RS-GIM (see for example \cite{RSGIMreview}).

\subsection{4-Fermi Operators}
$\Delta F = 2$ FCNCs are mediated by the 4-fermi operators $C_{1,4,5}$. For the up sector (identically for the down sector), the relevant terms in the effective Lagrangian are given by
\begin{eqnarray}
\label{e.4fermitotal}
\nonumber H^{(u)} &=& C_{1L\mathrm{(u)}}^{\alpha \beta \sigma \lambda} (\overline{q}_{uL}^\alpha \gamma_\mu {q}_{uL}^\beta) (\overline{q}_{uL}^\sigma \gamma^\mu q_{uL}^\lambda) + \{ L \rightarrow R \} + \\
\nonumber &&  C_{4\mathrm{(u)}}^{\alpha \beta \sigma \lambda} (\overline{q}^\alpha_{uR} q^\beta_{uL})(\overline{q}^\sigma_{uL} q^\lambda_{uR})
+ C_{5 \mathrm{(u)}}^{\alpha \beta \sigma \lambda} (\overline{q}^{c \alpha}_{uR} q^{d \beta}_{uL})(\overline{q}^{d \sigma}_{uL} q^{c \lambda}_{uR}),
\end{eqnarray}
where greek letters denote flavor indices and $c,d$ are color indices (if not shown, then color is contracted inside brackets).  We will compute the FCNC operators by integrating out the massive gauge bosons, then compare them to UTfit bounds \cite{UTfit}. The most relevant constraints come from meson mixing processes, i.e. $D$ mixing in the up sector and $K$, $B_d$ and $B_s$ mixing in the down sector:
\begin{equation}
C^D_{1L} = C_{1L\mathrm{(u)}}^{1212} \ \ \ \ \ \ \
C^K_{1L} = C_{1L\mathrm{(d)}}^{1212} \ \ \ \ \ \ \
C^{B_d}_{1L} = C_{1L\mathrm{(d)}}^{1313} \ \ \ \ \ \ \
C^{B_s}_{1L} = C_{1L\mathrm{(d)}}^{2323}
\end{equation}
(similarly for $C_{1R}, C_4, C_5$). Integrating out the massive gauge bosons in our model, we obtain:
\begin{align}
\label{e.4fermiops}
\nonumber C_{1L\mathrm{(u)}}^{\alpha \beta \alpha \beta} =& - \frac{1}{3}  \sum_{\mathrm{KK}} \frac{1}{m_G^2}\left({g^{\alpha \beta}_{G u_L u_L}}\right)^2 + \frac{1}{2}  \sum_{\mathrm{KK}} \frac{1}{m_Z^2}\left({g^{\alpha \beta}_{Z u_L u_L}}\right)^2\\
\nonumber C_{4\mathrm{(u)}}^{\alpha \beta \alpha \beta} =& -\sum_{\mathrm{KK}} \frac{1}{m_G^2} g^{\alpha \beta}_{G u_R u_R} g^{\alpha \beta}_{G u_L u_L}  - 2 \sum_{\mathrm{KK}} \frac{1}{m_Z^2}{g^{\alpha \beta}_{Z u_R u_R}}{g^{\alpha \beta}_{Z u_L u_L}}\\
C_{5\mathrm{(u)}}^{\alpha \beta \sigma \lambda} =& -\frac{1}{3} \sum_{\mathrm{KK}} \frac{1}{m_G^2} {g^{\alpha \beta}_{G u_R u_R}} {g^{\alpha \beta}_{G u_L u_L}}
\end{align}
where "KK" indicates that we sum over gauge KK modes, including the SM $Z$ boson. For each operator, we define a suppression scale $\Lambda$ by $|C| = \frac{1}{\Lambda^2}$.

\subsection{$G'$ contributions to FCNCs}
We want to write down expressions for the suppression scales of the $C_{1,4,5}$ operators due to contributions of the first gluon KK-mode, which dominate if $Z$-couplings are matched to the SM and (almost) flavor universal. The mass of the first KK-gluon is given to $10\%$ accuracy by
\begin{equation}
m_{G'} \approx \frac{x_1}{R'} \approx x_1 \sqrt L M_W
\end{equation}
where we have used \eref{Wmass}, and $x_1 \approx 2.4$ is the first root of the bessel function $J_0(x_1) = 0$. The bulk part of its coupling to a left-handed zero mode is approximately
\begin{equation}
\label{e.gluonKKcoupling}
g_L^\mathrm{bulk}  \approx g_{S*}  f(c_i)^2 \,\gamma(c_i) \equiv g_{S*} F(c),
\end{equation}
where $\gamma(c) =  \frac{\sqrt{2}}{J_1(x_1)} \int_0^1 x^{1-2c} J_1(x_1 x) dx
\approx \frac{\sqrt{2}}{J_1(x_1)} \frac{0.7}{6 - 4c} (1 + e^{c/2})$ is an $O(1)$ numerical correction factor \cite{RSGIMreview}\footnote{The accuracy of the approximate expression for $\gamma(c)$ in \cite{RSGIMreview} is somewhat improved by replacing $x_1 \rightarrow (1+e^{c/2})$}. Gauge matching sets $g_{S*} = \sqrt{4 \pi \alpha_S L}$, and we find that numerically, $m_{G'}/g_{S*} \approx 2 M_W$.

Now put everything together by substituting \eref{gluonKKcoupling} into \eref{offdiagNC}, and using those couplings in \eref{4fermiops}. We obtain the following expressions for the down-sector flavor suppression scales:
\begin{align}
\label{e.supscale}
\nonumber \Lambda_{1L}^{(d)ij} &\approx \dfrac{2 \sqrt{3} M_W}{U^{ij}_{Ld}} \left|F(c_{Q_L}) - F(c_L)\right|^{-1}\\
\nonumber \Lambda_{1R}^{(d)ij} &\approx \dfrac{2 \sqrt{3} M_W}{U^{ij}_{Ld}} \left|m_d^i m_d^j \left[\frac{F(-c_{Q_R})}{m_c^2 \rho_c^2} - \frac{F(-c_{b_R})}{m_b^2 \rho_b^2}\right]\right|^{-1}\\
\nonumber \Lambda_4^{(d)ij} &\approx \dfrac{2 M_W}{U^{ij}_{Ld}}\left|m_d^i m_d^j \left[F(c_{Q_L}) - F(c_L)\right]\left[\frac{F(-c_{Q_R})}{m_c^2 \rho_c^2} - \frac{F(-c_{b_R})}{m_b^2 \rho_b^2}\right]\right|^{-1/2}\\
 \Lambda_5^{(d)ij} &= \sqrt{3} \Lambda_4^{(d)ij}
\end{align}
where we replace $U^{21}_{Ld} \rightarrow U^{31}_{Ld} U^{32}_{Ld}$. For the up sector just change subscripts $d \rightarrow u$ and replace $\rho_b m_b, c_{b_R}$ by $\rho_t m_t, c_{t_R}$.

To see that our model has sufficient flavor protection to satisfy FCNC constraints we plug in some typical numbers and compare them to the RS-GIM suppression and the UTfit experimental flavor bounds on BSM FCNC contributions \cite{UTfit}\footnote{We evolve the UTfit bounds down to the KK scale using expressions in \cite{QCDRGE}. We thank Andreas Weiler for supplying the necessary code.}. The results are shown in \tref{allFCNC} and demonstrate why we need all our suppression mechanisms. The RS-GIM-like mechanism for the right-handed couplings, together with the flavor symmetry, ensures that the $C_4$ and $C_5$ operators are easily below bounds, a great improvement on traditional RS-GIM alone. The suppression scales of the $C_1$ operators are set by $\Lambda_{1L} \ll \Lambda_{1R}$, which are only Cabibbo suppressed, with the direct 12 contribution forbidden by the $SU(2)$ flavor symmetry. This is another reason why we need the flavor symmetry -- breaking it would increase 12-mixing by $\sim \lambda^4 \sim 500$, immediately violating bounds. Even with the flavor symmetry, the $C_1$ operators are close to bounds and the greatest source of angle constraints -- indeed, we can see that most of the mixing will have to be in the up-sector.

\begin{table}
\begin{center}
\begin{tabular}{@{}l|ll|ll}
Parameter & $\Lambda_F^\mathrm{bound}(3 \tev)$& RS-GIM $\Lambda_F$ & $\Lambda_F^\mathrm{bound}(0.7 \tev)$ & NMFV $\Lambda_F$\\
\hline \hline
Re$C_K^{1}$  & $1.0 \cdot 10^{3}$ & $\sim r /( \sqrt{6} \, |V_{td} V_{ts}| f_{q_3}^2) = 23 \cdot 10^{3}$& $1.1 \cdot 10^3$  &   $ 44 \cdot 10^3$\\
Re$C_K^{4}$  & $12 \cdot 10^{3}$ &  $\sim r (v Y_*)/(\sqrt{2\, m_d m_s})=22 \cdot 10^{3} $              & $11 \cdot 10^3$  & $19000 \cdot 10^3$\\
Re$C_K^{5}$  & $10 \cdot 10^{3}$ &  $\sim r (v Y_*)/(\sqrt{6\, m_d m_s}) =38 \cdot 10^{3}$              & $10 \cdot 10^3$ & $33000 \cdot 10^3$\\
\hline
Im$C_K^{1}$  & $16 \cdot 10^{3}$ & $\sim r/(\sqrt{6}\, |V_{td} V_{ts}| f_{q_3}^2) = 23 \cdot 10^{3}$    & $17 \cdot 10^3$   &$44 \cdot 10^3$\\
Im$C_K^{4}$  & $162 \cdot 10^{3}$ &$\sim r (v Y_*)/(\sqrt{2\, m_d m_s})=22 \cdot 10^{3} $               & $150 \cdot 10^3$  &$19000  \cdot 10^3$\\
Im$C_K^{5}$  & $147 \cdot 10^{3}$ &$\sim r (v Y_*)/(\sqrt{6\, m_d m_s}) =38 \cdot 10^{3} $              & $150 \cdot 10^3$  & $33000 \cdot 10^3$\\
\hline
\hline
$|C_{D}^{1}|$ & $1.3 \cdot 10^{3}$ & $\sim r/(\sqrt{6}\, |V_{ub} V_{cb}| f_{q_3}^2) = 25 \cdot 10^{3}$  & $1.3 \cdot 10^3$  &$1.8\cdot 10^3$\\
$|C_{D}^{4}|$ & $3.7 \cdot 10^{3}$ & $\sim r (v Y_*)/(\sqrt{2\, m_u m_c})=12 \cdot 10^{3} $             & $3.5 \cdot 10^3$  &$200 \cdot 10^3$\\
$|C_{D}^{5}|$ & $1.4 \cdot 10^{3}$ & $\sim r (v Y_*)/(\sqrt{6\, m_u m_c}) =21 \cdot 10^{3} $            & $1.5 \cdot 10^3$  &$500 \cdot 10^3$\\
\hline
\hline
$|C_{B_d}^{1}|$  & $0.22 \cdot 10^{3}$& $\sim r/(\sqrt{6}\, |V_{tb} V_{td}| f_{q_3}^2) = 1.2 \cdot 10^{3}$ & $0.22 \cdot 10^3$   &$0.35 \cdot 10^3$\\
$|C_{B_d}^{4}|$  & $1.7 \cdot 10^{3}$ & $\sim r (v Y_*)/(\sqrt{2\, m_b m_d})=3.1 \cdot 10^{3} $            & $1.6 \cdot 10^3$   &$24 \cdot 10^3$\\
$|C_{B_d}^{5}|$  & $1.3 \cdot 10^{3}$ & $\sim r (v Y_*)/(\sqrt{6\, m_b m_d}) =5.4 \cdot 10^{3} $           & $1.4 \cdot 10^3$   &$41 \cdot 10^3$\\
\hline
\hline
$|C_{B_s}^{1}|$ & $31$  & $\sim r/(\sqrt{6}\, |V_{tb} V_{ts}| f_{q_3}^2) = 270$ & $31$  &$70$ \\
$|C_{B_s}^{4}|$  & $210$ &$\sim r (v Y_*)/(\sqrt{2\, m_b m_s})= 780$            & $190$  &$1000$\\
$|C_{B_s}^{5}|$  & $150$ &$\sim r (v Y_*)/(\sqrt{6\, m_b m_s}) = 1400$          & $155$  &$1800$\\
\hline
\hline
\end{tabular}
\end{center}
\caption{We compare lower bounds\vspace{0.1mm} on the NP flavor scale $\Lambda_F$ (all in TeV) for arbitrary NP flavor structure from the UTFit collaboration \cite{UTfit} to the effective suppression scale in RS-GIM \cite{RSGIMreview} and our higgsless NMFV model, see \eref{supscale}. In this RS-GIM model, $|Y_*| \sim 3$, $f_{q_3}=0.3$ and $r = m_G/g_{s*}$, with a KK scale of $\sim 3 \tev$. For the higgsless model $L \approx 13$ determines a KK scale $m_G \approx 700 \gev$. Setting $\rho_c = 10$ gives $M_D = 110 \gev \sim 1/R'$, and $(c_{Q_L}, c_{Q_R}) = (0.48, -0.44)$ matches the couplings for the first two generations to the SM. A third generation EWPD match is most easily obtained for $\rho_{b,t} = 1$ and $(c_L, c_{b_R}, c_{t_R}) = (0.1, -0.73, 0)$. To satisfy the flavor bounds, we need to push more mixing into the up-sector by setting $ \lambda^{-1} U_{L(d)}^{13} \sim U_{L(u)}^{13} \sim\lambda^3$ and $\lambda^{-1} U_{L(d)}^{32} \sim U_{L(u)}^{32} \sim\lambda^2$.
}
\label{t.allFCNC}
\end{table}

\subsection{$Z$ contribution to FCNCs}
We have not explicitly estimated FCNC contributions due to $Z$-exchange, however they are included in the numerical scans in \sref{numerical}. They are negligible for the down sector, since all three diagonal couplings are matched to the SM, but the top coupling deviates by $O(40\%)$ in the full calculation, generating off-diagonal terms in the up-sector. The scale of the $Z'$ contributions to FCNCs (more important for LH than RH couplings, since $g_{Z u_\ell u_\ell} \approx 2 g_{Z u_r u_R}$) can be  estimated using $C^1 \sim (g_L/m_\mathrm{KK})^2$ and compared to the gluon KK contribution:
\begin{equation}
 O(0.4) \frac{g_{Zt_\ell t_\ell}^\mathrm{SM}}{m_Z}  \sim \frac{1}{10^3 \gev} - \frac{1}{10^2 \gev} \phantom{spacer} \mbox{and} \phantom{spacer}
\frac{g_{S*} F(c_L)}{m_G}  \sim \frac{1}{10^2 \gev}.
\end{equation}
Indeed, numerical scans in \sref{numerical} show that $Z$ contribution are negligible for all FCNC operators except $C_D^1$, where it does not invalidate the suppression mechanism but does supply a competitive contribution. In comparing FCNC's to experiment, one might worry that one has to take into account that the $Z$ contributes to FCNCs at a much lower scale than the KK modes. This is unnecessary, since the $C_D^1$ operator only changes by a few percent as we evolve it from our KK scale to the weak scale.

\subsection{Contribution to FCNCs from Higher-Dimensional Operators in the 5D Action}
Since 5D gauge theories are not renormalizable, our fermion action could include terms of the form
\begin{equation}
\int d^5x \sqrt{g} \frac{\Psi \overline\Psi \Psi \overline\Psi}{\Lambda^3},
\end{equation}
where $\Lambda = 16 \pi/g_5^2 = \Lambda_\mathrm{cutoff} R'/R $ (see eqn. \ref{e.cutoff}) is the unwarped 5D cutoff. The $SU(2)$ flavor symmetry forbids contributions of this form to 12 mixing, but they do contribute for 13 and 23 mixing\footnote{We thank Andreas Weiler for pointing this out.}. Since the right-handed quarks live almost entirely on the UV brane, where the cutoff is very high, we only have to worry about the left-handed quarks. The contribution to $C^1_{B_s}$ and $C^1_{B_d}$ is
\begin{equation}
\sim \frac{1}{\Lambda_\mathrm{cutoff}^3} \int dz \left( \frac{R}{z}\right)^5 \left( \frac{R}{R'}\right) (g_{b_L} g_{d_L})^2 \sim \frac{1}{(200 - 500 \tev)^2},
\end{equation}
depending on fermion localization. Comparing this to the experimental bounds in \tref{allFCNC} of $31$ and $22 \tev$ respectively, it is clear that we can ignore contributions by these operators.

%%%%%%%%%%%%%%%%%%%%%%%%%%%%%%%%%%%%%%%%%%%%%%%%%%%%%%%%%%%%%%%%%%%%%%
%%%%%%%%%%%%%%%%%%%%%%%%%%%%%%%%%%%%%%%%%%%%%%%%%%%%%%%%%%%%%%%%%%%%%%%
\section{Numerical Results and Mixing Constraints}
\label{s.numerical} \setcounter{equation}{0} \setcounter{footnote}{0}
%%%%%%%%%%%%%%%%%%%%%%%%%%%%%%%%%%%%%%%%%%%%%%%%%%%%%%%%%%%%%%%%%%%%%%
%%%%%%%%%%%%%%%%%%%%%%%%%%%%%%%%%%%%%%%%%%%%%%%%%%%%%%%%%%%%%%%%%%%%%%
We will now perform numerical scans to verify the results of the zero mode calculation and explicitly demonstrate that the higgsless NMFV model can satisfy flavor constraints. This is necessary because, once we have chosen our gauge sector, fermion bulk masses and IR brane terms, there is an overall rotation amongst the UV kinetic terms that is unconstrained by electroweak precision data and determines the FCNCs.

Our method for this scan is as follows: we will first perform some calculations without flavor mixing, which incorporate the diagonal brane terms into the boundary conditions to capture all KK mixing effects and match the SM couplings. Assuming any small flavor mixing would not change the diagonal couplings by much, we can use these calculations as a guideline in choosing our bulk and IR masses for a fully mixed calculation. We then explicitly calculate FCNCs for those input parameters by scanning over allowed down-sector mixing angles. This initial scan will be performed in the zero mode calculation for computational efficiency. Since there will likely be sizeable errors in the up-sector, we will take those points which passed FCNC bounds and recalculate them with full KK- and T-mixing, discarding those which now lie beyond bounds.

We should note that \emph{exact} compliance with EWPD is not required for this scan, since a small adjustment to the input parameters (to correct any small deviations) would not change the FCNCs significantly. At any rate, using the zero mode calculation to match flavor rotations introduces order unity errors into the charged-current mixing angles, which would have to be corrected by readjusting the up-sector rotations. We can do without such complications, since we only strive for Cabibbo-\emph{type} mixing in our scan, and most of the flavor constraints are in the down sector. If our scan indicates that FCNCs are under control for a general Cabibbo-type mixing, then they should also be under control for an exact CKM match.

\subsection{Input Parameters}
For our gauge sector, we choose $g_{5L} = g_{5R}$, $R^{-1} = 10^8 \gev$ and set effective BKTs at the weak scale to zero or as small as possible. This gives $L \approx 13$ and the highest possible cutoff scale.

In order to match the bottom couplings, we set both $M_1$ and $M_3$ to their minimum values and move only the $b_R$ close to the UV. From an unmixed calculation with full BCs we find the following values:
\begin{equation}
c_{b_R} = -0.73, \ \ \ \  \ c_{t_R} = 0, \ \ \ \ \ c_L = 0.1, \ \ \ \ \ M_1 = 600 \gev, \ \ \ \ \ M_3 = 140 \gev.
\end{equation}

We also know that we need to ramp up $M_D$ beyond its minimum value to satisfy constraints on the $C_{4,5}$ operators, so we choose $\rho_c = 10$. In order to pick bulk masses for the first two generations, we run another unmixed calculation with full boundary conditions and select three possible $(c_{Q_L}, c_{Q_R})$ values to run angle scans for:
\begin{center}
\begin{tabular}{l|rrr}
Scan  & 1 & 2 & 3\\
\hline
$c_{Q_R}$ & $-0.37$ & $-0.44$ & $-0.57$\\
$c_{Q_L}$ & $0.48$ & $0.48$ & $0.57$\\
$M_D$ $(\gev)$ & 76 & 101 & 445
\end{tabular}
\end{center}

\subsection{Angle Scans}
We will calculate FCNCs due to tree-level $Z, Z', G', G''$-exchange, for one million different down-sector mixings per scan. We can parametrize the CKM mixing matrix as
\begin{equation}
V(s_{12}, s_{23}, s_{13}, \delta) = \left(
\begin{array}{ccc}
 c_{12} c_{13} & c_{13} s_{12} & s_{13} e^{-i \delta} \\
 -c_{23} s_{12}- c_{12} s_{13} s_{23}e^{-i \delta}  & c_{12} c_{23}-   s_{12} s_{13} s_{23} e^{-i \delta} & c_{13} s_{23} \\
 -  c_{12} c_{23} s_{13}e^{-i \delta} +s_{12} s_{23} & - c_{23} s_{12} s_{13}e^{-i \delta} -c_{12} s_{23} & c_{13} c_{23}
\end{array}
\right)
\end{equation}
where, for example, $\left(s_{12}, s_{23}, s_{13}, \delta\right)_\mathrm{CKM} = \left( 0.227, 0.0425, 4\cdot 10^{-3}, 0.939\right)$ would satisfy the PDG constraints on $V_\mathrm{CKM}$ \cite{PDG}. Naively, we would think that we can obtain the correct CKM matrix by defining our up- and down-sector LH rotations as
\begin{equation}
\label{e.ULintermsofU}
U_{Lu} = U V_\mathrm{CKM}^\dagger \ \ \ \ \ U_{Ld} = U,
\end{equation}
and letting $U$ be an arbitrary unitary rotation matrix which gives the down-sector mixing. This is sufficient for this scan, even though it only gives an order unity estimate of the up-sector mixing angles. In this analysis we shall also ignore phases, since we are after a scan of the \emph{magnitudes} of the possible mixing matrices, and for our purposes we define $V_\mathrm{CKM} = V(s_{12}^\mathrm{CKM}, s_{23}^\mathrm{CKM}, s_{13}^\mathrm{CKM}, 0)$ (otherwise we could never cancel this matrix with a real rotation, introducing an up-mixing bias into our scan). To avoid obviously large FCNCs, we will make the assumption that the mixing angles of $U$ have a \emph{natural size} comparable to those of the final $V_\mathrm{CKM}$ mixing matrix. We parametrize $U$ with angle-coordinates $(a,b,c) \sim O(1)$:
\begin{equation}
\label{e.Uangledef}
U = V(s_{12}, s_{23}, s_{13}, 0) \ \ \ \mbox{where} \ \ \  s_{12} = a s_{12}^\mathrm{CKM} \ \ \ s_{23} = b s_{23}^\mathrm{CKM} \ \ \ s_{13} = c s_{13}^\mathrm{CKM}
\end{equation}
Note that $(a,b,c) = (0,0,0)$ and $(a,b,c) = (1,1,1)$ put all the mixing into the up- and down-sector respectively, so to avoid a bias in our scan we define the range of the angle coordinates to be $a,b,c \in (-2,3)$. Once we determine which points in angle-space satisfy FCNCs in the zero mode calculation, we re-check those points using a full KK calculation.

\subsection{Results}
As we can see from the similar plots in \fref{scanfit}, the choice of bulk masses does not have a great effect on the nature of constraints on the down-mixing angles. This is expected since the $C_1$ operators, which are the greatest bottleneck, are only weakly dependent on $c_{Q_R}, c_{Q_L}$ -- the dominant contribution comes from the large $F(c_L)$, see \eref{supscale}. In eliminating points which do not satisfy FCNC bounds with full KK mixing, we only loose a few percent of points in each scan. The zero mode calculation is therefore sufficient for estimating the angle constraints.

\begin{figure}
\begin{center}
\includegraphics[keepaspectratio, width = 16cm]{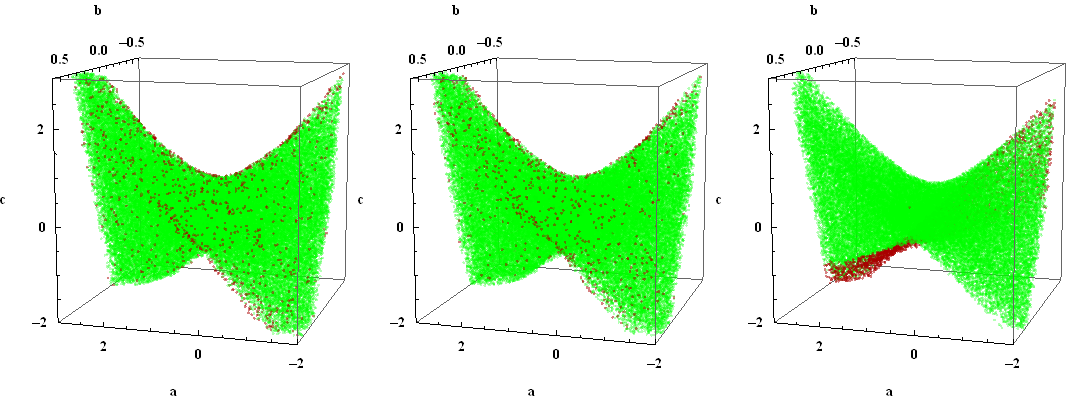}
\caption{Left to right: Points in $U_{Ld}$ angle space for scans 1, 2 and 3 that satisfy FCNC constraints in the zero mode calculation, where $s_{12} = a s_{12}^\mathrm{CKM}, s_{23} = b s_{23}^\mathrm{CKM}$ and $s_{13} = c s_{13}^\mathrm{CKM}$. Dark red points are found to violate the bounds when taking into account KK mixing.}\label{f.scanfit}
\end{center}
\end{figure}

\begin{figure}
\begin{center}
\includegraphics[keepaspectratio, width = 16cm]{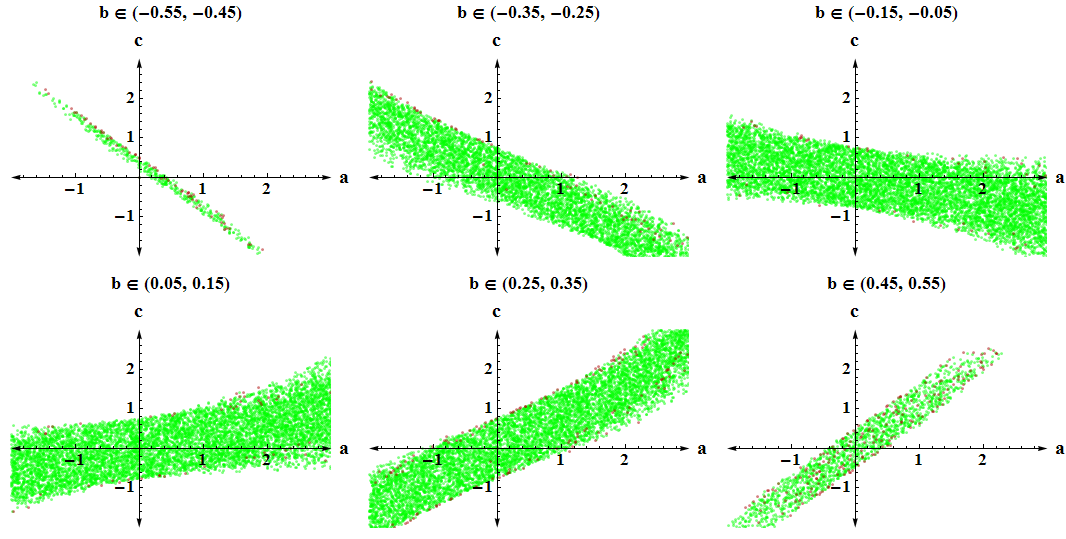}
\caption{Slices of thickness $0.1$ in the $(a,c)$-plane at different points along the $b$-axis, from $b = -0.5$ to $b = 0.5$, for scan 2 (scans 1 and 3 are similar). Green (light) points satisfy FCNC constraints in the zero mode calculation, and red (dark) points fail bounds when KK mixing is taken into account. The points which fail FCNC bounds in the zero mode calculation are not shown, but they fill up the entire remaining volume of angle space. There is \emph{no overlap} between points satisfying FCNC bounds and points that do not -- they occupy well defined, mutually exclusive volumes.}\label{f.scanfitresult44bslices}
\end{center}
\end{figure}

The FCNC bounds impose entirely systematic constraints on the down-sector mixing angles. This can be seen from \fref{scanfitresult44bslices}, where we take slices at different points on the $b$-axis (i.e. $s_{23}$) and project them onto an $ac$-plane.  A point in angle-space satisfies FCNC bounds if and only if it lies within a well-defined sub-volume, i.e. the constraints are systematic. Assuming Cabibbo-type mixing, the good points occupy $\sim O(5 \%)$ of the total angle space. This is not really `tuning' in the usual sense, it merely means that whatever UV-scale mechanism generates the mixings should give a somewhat larger mixing in the up-sector than in the down-sector. We note that while $s_{12}$ and $s_{13}$ are correlated, their range is fairly unconstrained, whereas $s_{23}$ must fall within strict limits to satisfy $C^1_{B_s}$  constraints. Roughly speaking, less than half the 23-mixing is allowed to be in the down sector.

We can conclude that our higgsless NMFV model should have no trouble satisfying FCNC bounds as long as certain systematic constraints on the down-sector mixing angles are met.

%%%%%%%%%%%%%%%%%%%%%%%%%%%%%%%%%%%%%%%%%%%%%%%%%%%%%%%%%%%%%%%%%%%%%%
%%%%%%%%%%%%%%%%%%%%%%%%%%%%%%%%%%%%%%%%%%%%%%%%%%%%%%%%%%%%%%%%%%%%%%%
\section{Conclusion}\vspace{-2mm}
\label{s.conclusion} \setcounter{equation}{0} \setcounter{footnote}{0}
%%%%%%%%%%%%%%%%%%%%%%%%%%%%%%%%%%%%%%%%%%%%%%%%%%%%%%%%%%%%%%%%%%%%%%
%%%%%%%%%%%%%%%%%%%%%%%%%%%%%%%%%%%%%%%%%%%%%%%%%%%%%%%%%%%%%%%%%%%%%%
We examined various possibilities for higgsless RS model-building, and constructed a model with next-to-minimal flavor violation satisfying tree-level electroweak precision and meson-mixing constraints, as well as CDF bounds. The theory has a sufficiently high cutoff of $\sim 8 \tev$ to unitarize $WW$-scattering at LHC energies, the third generation is in the custodial quark representation to protect the bottom couplings, and a combination of flavor symmetries and UV confinement of the right-handed quarks suppress FCNCs. Using numerical scans, we were able to demonstrate that our model can satisfy flavor bounds as long as the down-sector mixing angles are Cabibbo-type and satisfy systematic constraints. We also found quantitative error estimates for the zero mode approximation, which are important for RS model-building with a low KK scale.

This model has distinctive experimental signatures, allowing it to be excluded early on at the LHC. Apart from the absence of the Higgs, the usual higgsless RS signals include~\cite{higgslesscollider} a relatively light $G'$ with a mass below 1 TeV, as well as $Z'$ and $W'$ which are harder to detect (see Section \ref{ss.CDF}). More specific to our setup is an exotic $X$-quark with charge $5/3$ and a mass of $\sim 0.5 \tev$, which could be detected with less than $100$ pb$^{-1}$ of data \cite{Xquark}.  The NMFV model also predicts non-zero correlated flavor-changing neutral currents, which lie relatively close to current experimental bounds and would be detected in the next generation of flavor experiments.

%%%%%%%%%%%%%%%%%%%%%%%%%%%%%%%%%%%%%%%%%%%%%%%%%%%%%%%%%%%%%%%%%%%%%%
%%%%%%%%%%%%%%%%%%%%%%%%%%%%%%%%%%%%%%%%%%%%%%%%%%%%%%%%%%%%%%%%%%%%%%%
\section*{Acknowledgements} \vspace{-2mm}
\setcounter{equation}{0} \setcounter{footnote}{0}
%%%%%%%%%%%%%%%%%%%%%%%%%%%%%%%%%%%%%%%%%%%%%%%%%%%%%%%%%%%%%%%%%%%%%%
%%%%%%%%%%%%%%%%%%%%%%%%%%%%%%%%%%%%%%%%%%%%%%%%%%%%%%%%%%%%%%%%%%%%%%
We are grateful to Andreas Weiler, Kaustubh Agashe and Yuval Grossman for many useful discussions and comments on the manuscript, and in particular we thank Andreas Weiler for his help with calculating the RGE evolution of the UTfit flavor bounds. This research has been supported in part by the NSF grant number NSF-PHY-0757868. C.C. was also supported in part by a U.S.-Israeli BSF grant.

%%%%%%%%%%%%%%%%%%%%%%%%%%%%%%%%%%%%%%%%%%%%%%%%%%%%%%%%%%%%%%%%%%%%%%
%%%%%%%%%%%%%%%%%%%%%%%%%%%%%%%%%%%%%%%%%%%%%%%%%%%%%%%%%%%%%%%%%%%%%%%
%%%%%%%%%%%%%%%%%%%%%%%%%%%%%%%%%%%%%%%%%%%%%%%%%%%%%%%%%%%%%%%%%%%%%%
%%%%%%%%%%%%%%%%%%%%%%%%%%%%%%%%%%%%%%%%%%%%%%%%%%%%%%%%%%%%%%%%%%%%%%%

\end{document}